\documentclass[aps,prl,reprint,superscriptaddress]{revtex4-1}

\usepackage{graphicx}
\usepackage{mathtools}
\usepackage{amsmath}
\usepackage{amsfonts}
\usepackage{xcolor}
\usepackage{dsfont}
\usepackage{hyperref}
\usepackage[top=1in,left=0.5in,right=0.5in]{geometry}
\usepackage{floatrow}

\def\be{\begin{equation}}
\def\ee{\end{equation}}
\def\p{\partial}
\def\til{\tilde}
\def\f{\frac}

\def\l{\left}
\def\r{\right}
\def\<{\langle}
\def\>{\rangle}

\def\lam{\lambda}
\def\th{\theta}

\def\del{\delta}
\def\Del{\Delta}
\def\Order{\mathcal{O}}
\newcommand{\rfeq}[1]{{eq{.\bf~\ref{#1}}}}
\newcommand{\rfequation}[1]{{equation \bf\ref{#1}}}
\newcommand{\eq}{\textnormal{eq}}
\newcommand{\bb}[1]{\boldsymbol{{#1}}}
\newcommand{\FE}{\mathcal{F}}
\newcommand{\EDiss}{E_{\textnormal{diss}}}
\newcommand{\SProd}{S}
\newcommand{\SProdRate}{\dot{S}}
\newcommand{\ThLen}{\mathcal{L}}
\newcommand{\al}{\alpha}
\newcommand{\bt}{\beta}
\newcommand{\ga}{\gamma}

\def\d{\mathrm{d}}

\begin{document}

\title{Energy Dissipation Bounds for Autonomous Thermodynamic Cycles}
\date{\today}
\author{Samuel J. Bryant}
\email{samuel.bryant@yale.edu}
\affiliation{Department of Physics, Yale University, New Haven}
\author{Benjamin B. Machta}
\email{benjamin.machta@yale.edu}
\affiliation{Department of Physics, Yale University, New Haven}
\affiliation{Systems Biology Institute, Yale University, New Haven}


\begin{abstract}
How much free energy is irreversibly lost during a thermodynamic process?
For deterministic protocols, lower bounds on energy dissipation arise from the thermodynamic friction associated with pushing a system out of equilibrium in finite time.
Recent work has also bounded the cost of precisely moving a single degree of freedom.
Using stochastic thermodynamics, we compute the total energy cost of an autonomously controlled system by considering both thermodynamic friction and the entropic cost of precisely directing a single control parameter.
Our result suggests a challenge to the usual understanding of the adiabatic limit: here, even infinitely slow protocols are energetically irreversible.
\end{abstract}

\maketitle

\begin{figure}
\centering
\includegraphics{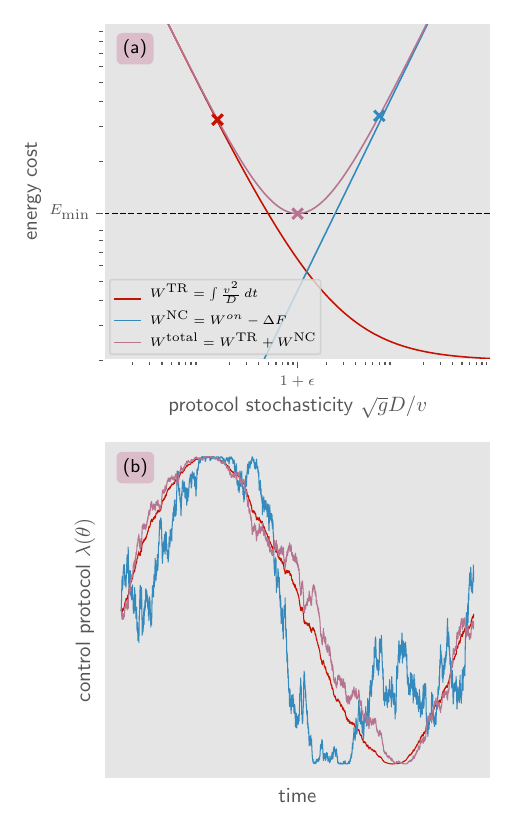}
\caption{(a) The contribution to the energy dissipation from the nonconservative work done on the system (blue), the contribution from the cost of directing the control protocol in time (red), and the total energy dissipation cost (purple).
On the $x$ axis is the ratio of the diffusion constant $D$ of the control protocol to the the speed $v$ of the control protocol, which captures the level of stochasticity of the protocol.
The marked points in (a) correspond to different levels of stochasticity in the control protocol.
For each mark, a sample trajectory with that stochasticity is shown in (b).
The blue mark is a suboptimal protocol which is too stochastic, doing more work than is necessary on the system.
The red mark is a suboptimal protocol which is too deterministic, spending more energy on reducing noise than is necessary.
The purple mark is an optimal protocol which minimizes this energetic tradeoff by balancing these two constraints.
}
\label{fig:energy_curves}
\end{figure}

Controlling the state of a thermodynamic system requires interacting with and doing work on its degrees of freedom.  
Given an initial thermodynamic state $\lam_i$ and a final state $\lam_f$, the second law of thermodynamics implies that the average work required to produce the transition between these states is at least the difference in their free energy: $\<W\>\geq F(\lam_f)-F(\lam_i)$.
Recent work has strengthened this inequality, quantifying two distinct sources of energetic dissipation.
First, in order to transform a system at a finite rate, energy must be expended to overcome thermodynamic friction~\cite{Jarzynski97,SivakCrooks12B,SivakCrooks12A,LargeSivak18,BurbeaRao82,Crooks99,Kirkwood46}.
Second, energy must be expended to bias the motion of $\lambda$ itself in a preferred direction~\cite{CaoTu15,FengCrooks08,BrownSivak17,ParrondoKawai09,Costa05,zhang_energy_2019,Barato16Cost,Verley14The}.
Here we consider both of these energetic costs together.
We find that while either cost can be made arbitrarily small, they cannot be minimized simultaneously, as illustrated in Figure~\ref{fig:energy_curves}.
This leads to a lower bound on the required energetic cost for controlling a thermodynamic system that remains non-zero {\it even in the limit that the process moves infinitely slowly and imprecisely.}

Any process that transforms a system in a {\it finite} amount of time must necessarily push the system out of equilibrium.
This requires doing non-conservative work which causes the irreversible loss of usable energy, a familiar example of which is the heat dissipated while moving an object through a viscous medium.
This dissipative cost, incurred by thermodynamic friction, can be written as: $\<\EDiss\>\equiv \<\Delta S\> = \<W\> - \Delta F$, where the dissipation equals $\Delta S$, the change in the entropy of the system and environment\footnote{
Since $\Delta F=\Delta E-T\Delta S$ and since the energy of the universe $E$ is constant, changes in the total amount of free energy can be identified with changes in the entropy of the universe.
}~\cite{Seifert05}.
Recent work has explicitly quantified this form of dissipation, finding an elegant geometric interpretation for the non-conservative work done in a finite time process~\cite{SivakCrooks12A,SivakCrooks12B}.

In the above analysis, the energetic cost is identified with $\<W\>$, the average work done on the thermodynamic system by an externally defined and deterministic control protocol $\lam(t)$.
However, for an autonomous system which implements its own control protocol, energy must be dissipated in order to move $\lam$ itself in a directed manner.
Physical systems obey microscopically time-reversible dynamics~\footnote{Though not all physical process are fully invariant under time reversal, the distinction is not important for out current understanding of how this type of directionality arises in in biological end engineered systems.}.
Thus to bias the motion of $\lam$, breaking time-reversal symmetry, forward progress must be accompanied by an increase in entropy, which can be accomplished by dissipating energy into a heat bath.
This entropic increase causes forward movement to be more likely than backwards movement, and without this, there can be no ``arrow of time''~\cite{FengCrooks08}.
A classic example of this is the consumption of ATP by myosin motors to ``walk'' along actin polymers~\cite{Pollard73}.
Without the irreversible consumption of free energy, the system would be constrained to have equilibrium dynamics, transitioning backward and forward at the same rate.
Recently, there has been considerable work done in understanding the tradeoffs between this bias cost and the accuracy and speed of system trajectories~\cite{Verley14Work,Verley14Universal,Verley14The,CaoTu15,Barato16Cost,Barato17Coherence,PietzonkaSeifert16,zhang_energy_2019}, culminating with the development of so-called thermodynamic uncertainty relations~\cite{,Gingrich16,Barato15Thermodynamic,Horowitz17}.

Here we consider the total dissipative cost of autonomously controlled thermodynamic systems, including contributions from both thermodynamic friction and entropic bias.
In particular, we consider thermodynamic systems with state $\bb x$ and control parameter $\lambda$,
where the dynamics of $\lam$ are now {\it stochastic} due to the fact that the entropic bias required to run a deterministic path is infinite~\cite{Crooks99}.
Additionally, we suppose that the dynamics of $\bb x$ depend on $\lam$ but the dynamics of $\lam$ do not receive feedback from $\bb x$.
This one-way coupling distinguishes $\lam$ as a physical {\it control parameter}
rather than as simply a degree of freedom in a larger thermodynamic system.
We can think of $\lam$ as mediating a coupling between $\bb x$ and different thermodynamic baths, as in Ref.~\cite{Machta15}.
This setup is analogous to a control knob changing the amount of power flowing into a thermodynamic system in that the work required to turn the knob\footnote{
  Since the cost of turning a control knob doesn't scale with the system size, for any macroscopic system one would naturally ignore this cost.
} is independent of the state of the system.

Examples of this class of processes are found throughout biology in systems that undergo regulated change.
In these systems, energy is expended both in running the regulatory mechanism and in instantiating downstream changes~\cite{Goldbeter96,Pomerening03,RodenfelsHoward19}.
In muscles, the release of calcium modulates the pulling force by activating myosin heads which then hydrolyze ATP to produce a step~\cite{SzentGyorgyi75,EisenbergHill85}.
In this case, the bias cost is the energy used to control the amount of calcium that is released through concentration gradients.
Energy is also dissipated through the non-conservative work that the muscles themselves do as they consume ATP to contract.
Another example is biological cycles, where energy is used both to coordinate phase in the cycle and to run downstream processes.
Cyclins, which underlie the cell cycle and KaiABC proteins which implement circadian oscillations, are kinases that phosphorylate each other to implement a cycle as well as downstream targets to instantiate the changes that must occur throughout the cycle~\cite{ZonWolde07,NakajimaKondo05,Terauchi07,Monti18,Rust07,RodenfelsHoward19,zhang_energy_2019,zwicker_robust_2010}.
Phosphorylating these downstream targets also consumes energy.
In both of these examples, the control degrees of freedom (calcium concentration and cyclin phosphostate) receive minimal feedback from their downstream target's state (contraction state of myosin heads and phosphostate of targets).

Our central result is that thermodynamic processes cannot be controlled in an energetically reversible manner.
While each of these two energetic costs can be made small in isolation, they cannot be made small together.
The cost of thermodynamic friction can only be minimized by moving the system slowly and precisely, while the bias cost can only be made small by moving imprecisely and with large fluctuations.
These competing constraints cannot be satisfied at the same time (see Figure~\ref{fig:energy_curves}).
We thus suggest that the notion of an energetically reversible control process exists only as an abstraction for when the control itself is considered external to the object under study.
A realistic self-contained thermodynamic machine has no reversible transformations, even in the quasistatic limit.

Our work builds on a framework developed in Ref.~\cite{SivakCrooks12A,SivakCrooks12B} which considered lower bounds on the energy dissipation arising from a deterministic control protocol $\bb \lambda(t)$ moving at finite speed.
Those authors found that the dissipation rate is given by:
\be
  \label{eq:sivak_crooks_result}
  \<\SProdRate\>
  \approx
  \f{\d\lam^\al}{\d t}\til{g}^\lam_{\al\bt}\f{\d\lam^\bt}{\d t}
\ee
where $\til{g}^\lam$ is the Kirkwood friction tensor~\cite{Kirkwood46}
and where we have replaced the energy dissipation $\EDiss$ with the equivalent notion of entropy production~\cite{Seifert05,Seifert12}.
This rate yields a lower bound on the energy dissipation:
\be
  \label{eq:sivak_crooks_bound}
  \SProd \geq \f{\til{\ThLen}^2(\bb\lam_i,\bb\lam_f)}{\Del t}
\ee
where $\til{\ThLen}$ is the minimum length path in $\bb\lam$ space between $\bb\lam_i$ and $\bb\lam_j$ with the metric $\til{g}^\lam_{\al\bt}$~\cite{Crooks07}
and where $\Del t$ is the total time of the protocol.
Energetically optimal control protocols are geodesics in the Riemannian thermodynamic space defined by $\til{g}^\lam_{\al\bt}$.

Importantly, this result implies that in the quasistatic limit where the protocol moves infinitely slowly ($\Del t\to\infty$), the excess energy cost of transforming the system becomes negligible (see \rfeq{eq:sivak_crooks_bound}).
However, in this framework, the breaking of time-reversal symmetry arises from the deterministic trajectory $\bb \lam(t)$, whose cost is neglected from this energetic book-keeping.

When we include the cost of the time-reversal symmetry breaking in the control protocol $\bb\lambda(t)$, the optimal control protocols are no longer deterministic.
Instead, we consider a protocol $\bb\lambda(\th)$ where $\theta$ is a stochastic variable that moves with net drift velocity $v$ and diffusion $D$.
When we include the bias cost of the control protocol itself, we find that the total rate of energy dissipation in the near-equilibrium limit is given by:
\begin{align}
  \l\<\SProdRate\r\>_\th
  \approx
  \l\<\f{\d\lam^\al}{\d t^+}\r\>_{\hspace*{-4pt}\th}
  \til{g}^\lam_{\al\bt}
  \l\<\f{\d\lam^\al}{\d t^-}\r\>_{\hspace*{-4pt}\th}
  +
  \f{v^2}{D}
  +
  D g^\th
  \label{eq:mainresult}
\end{align}
where $g^\th$ is the Fisher information metric on $\th$ (see \rfeq{eq:fisher_information}).
The first term of \rfeq{eq:mainresult} captures the thermodynamic friction associated with pushing the system out of equilibrium.
We will see that this term is the natural generalization of \rfeq{eq:sivak_crooks_result} to a stochastic system.
The second term captures the dissipation associated with breaking time reversal symmetry, moving the control protocol forward in time.
The third term is the result of the redundant thermodynamic friction caused by the stochastic system trajectory.

From this rate, we show that for a simple system, the total energetic cost of moving the system from $\bb\lam_i$ to $\bb\lam_f$ has a lower bound given by:
\be
  S \geq 2\ThLen(\bb\lam_i, \bb\lam_f) + \f{\til{\ThLen}^2(\bb\lam_i,\bb\lam_f)}{\Del t}(1-\epsilon)
\ee
where $\ThLen$ and $\til{\ThLen}$ are the lengths traversed in thermodynamic space with metrics $\til{g}$ and $g$ and where $\epsilon$ is a small correction that disappears as $\Del t\to \infty$, or in the limit of large protocol size.

This result is significant because the first term defies the quasistatic limit, {\it remaining non-zero even in the limit of an infinitely slow protocol} ($\Del t\to\infty$).
The result we obtain from the simple system analyzed here suggests a more general principle: every thermodynamic transformation has a minimal energy requirement that cannot be made arbitrarily small by using a slow control protocol.
This energy requirement is determined by the geometry of thermodynamic space.

Our result generalizes and clarifies a bound anticipated in Ref.~\cite{Machta15}, which considered coupling a thermodynamic system to a succession of particle reservoirs.
Here we  explicitly take a continuum limit, and consider protocols which move at finite rate, connecting this to other bounds in the literature~\cite{SivakCrooks12A}.
Our results clarify that this bound is far more general, showing that the minimum energy bound is not specific to that class of systems but is rather a fundamental property of thermodynamic machines.
In particular, it holds for any thermodynamic system in which a single degree of freedom obeying the microscopic laws of thermodynamics mediates a coupling to a generalized thermodynamic reservoir.

\section*{Derivation}
Consider a system of particles interacting through an energy of the form:
\be
  U(\bb y) = \lam^\alpha(\th)\phi_\alpha(\bb x)
  \qquad
  \bb y = (\th, \bb x)
\ee
where $\bb x$ denotes the microstate of the system and $\th$ parametrizes the path of the control functions around a cycle.
Here we use greek letters $\alpha$,$\beta$,$\gamma$, etc to index the controlling potentials $\bb\lam$ and their conjugate variables $\bb\phi$
and we use roman letters $i$, $j$, $k$ to index the microstate $\bb x$.
We also use natural units and unit temperature such that $\bt=T=1$.

The control functions $\bb \lam(\th)$ are all $2\pi$ periodic in $\theta$,
allowing the system to reach a steady state where the system cycles are fully self-contained.
We suppose that the control parameter $\th$ is driven stochastically with net drift velocity $v$ and diffusion constant $D$.
We describe both $\th$ and $\bb x$ using overdamped Langevin equations~\cite{Seifert12}:
\be
\label{eq:langevin_equations}
\begin{array}{cc}
  \begin{aligned}
    \dot{\th} &= v + \sqrt{2D}\eta
    \\
    \dot{x}^i &= D^{ij} F_j(\bb y) + \sqrt{2D^{ij}}\xi_j
  \end{aligned}
  &
  \begin{aligned}
    \<\eta(t)\eta(t')\>&=\del(t-t')
    \\
    \<\xi_i(t)\xi_j(t')\>&=\del_{ij}\del(t-t')
  \end{aligned}
\end{array}
\ee
The terms $\eta$ and $\xi^i$ represent thermal noise.
The overdamped force on $x^i$ due to the interaction energy $U$ is given by $F_i(\bb y) = -\p_{x^i} U(\bb y)$.
As discussed in the introduction, by {\it control parameter} we mean precisely that the term $F_\th(\bb y) = -\p_\th U(\bb y)$ is absent from the stochastic equation for $\dot{\th}$;
the dynamics of $\th$ don't receive feedback from the state of the system.

The Langevin dynamics of the system give rise to probability currents:
\begin{align}
  j^\th \equiv \l[v-D\p_\th\r]&p(\bb y,t)
  &&
  j^i \equiv D^{ij}[F_j(\bb y) - \p_{x_i}] p(\bb y,t)
\end{align}
Using stochastic thermodynamics, we can derive the average rate of energy dissipation by equivalently calculating the average rate of total entropy production~\cite{Seifert12}:
\be
  \<\SProdRate(t)\> = \int \d\bb y\,\l[
  \f{[j^\th(\bb y,t)]^2}{D p(\bb y,t)}
  +
  \f{j^i(\bb y,t)[D^{-1}]_{ij}j^j(\bb y, t)}{p(\bb y,t)}
\r]
\ee

Because the system is periodic and the Langevin dynamics (\rfeq{eq:langevin_equations}) are time-independent, the system will eventually reach a steady-state that can be described by the joint probability distribution $p(\bb y)$.
To compute the energy dissipation, we first compute the average rate of energy dissipation conditioned on the the control parameter being in the known state $\th$.
We can then compute the total energy dissipation by averaging over a full cycle in $\th$.
Denote $p(\bb x|\th)$ to be the non-equilibrium steady-state probability of finding the system in the microstate $\bb x$ given that the system is at $\th$.
Denote $\<\cdots\>_{\th}$ to mean an average over the non-equilibrium distribution $p(\bb x|\th)$ for a fixed $\th$.
Finally, denote $\<\cdots\>_{\th,\eq}$ to mean an average over the equilibrium Boltzmann weight $p_{eq}(\bb x|\th)\propto e^{-U(\th,\bb x)}$.
We find the rate of entropy production when the system is at $\theta$ is given by (for the full details of the calculation see \rfeq{si:main-derivation-start}-\rfeq{si:main-derivation-end} in the Supplement):
\be
  \label{eq:dissipation_no_linear_response}
  \<\SProdRate\>_\th
  =
  \f{v^2}{D}
  +
  D g^\th(\th)
  +
  \<\del\phi_\alpha\>_{\th}
  \l[
    v
    \f{\p\lam}{\p\th}^\alpha\hspace*{-3pt}(\th)
    +
    D
    \f{\p^2\lam^\alpha}{\p\th^2}(\th)
  \r]
\ee
where $\<\del\phi_\alpha\>_\th=\<\phi_\alpha\>_\th-\<\phi_\alpha\>_{eq,\th}$ is the deviance of the conjugate force $\phi_\alpha$ at $\th$ from its equilibrium value and
and $g^\th$ is the Fisher information metric with respect to the $\th$ basis:
\be
  \label{eq:fisher_information}
  g^\th(\th)
  \equiv \f{\d\lam}{\d\th}^\alpha g^\lam_{\alpha\beta}(\lam)\f{\d\lam}{\d\th}^\beta
  \equiv
  \f{\d\lam}{\d\th}^\alpha\<\del\phi_\alpha\del\phi_\beta\>_{\eq,\lam}\f{\d\lam}{\d\th}^\beta
\ee
The Fisher information metric can be thought of as a measure of how strongly the distribution $p_\eq(\bb x|\th)$ changes as the value of $\th$ changes as determined by the Kullback-Leibler divergence.

So far this is an exact result.
In the next section we will use linear response to approximate $\<\del\phi_\alpha\>_\th$, assuming that $D$ and $v$ are small.

\subsection*{Linear Response for Stochastic Control}
Since we are interested in a lower energetic bound, we consider the low dissipation regime where the $\bb x$ system is driven near-equilibrium.
In this regime, we can use linear response to compute the thermodynamic friction of the $\bb x$ system.
To evaluate the term $\<\del\phi_\alpha\>_{\th}$, we use a linear response approximation.
For a fixed control parameter path $\th(\tau)$, the average linear response over all possible {\it microstate paths} is well understood~\cite{SivakCrooks12A}.
Here we extend this result to an average over all possible microstate {\it and} control parameter paths.

If an ensemble of systems all undergo the same {\it deterministic} protocol $\th(\tau)$, then at time $t$, the average linear response of $\phi_\alpha$ over this ensemble is given by~\cite{Zwanzig01}:
\be
  \label{eq:basic_linear_response}
  \<\del\phi_\alpha(t)\>_{\th(\tau)}
  =
  \int_{-\infty}^0\d t'\,
  C^{\th(t)}_{\alpha\beta}(t')
  \f{\d}{\d t'}\l[\lam^\beta(t+t')\r]
\ee
where $C^{\th}_{\alpha\beta}(t') = \<\del\phi_\alpha(0)\del\phi_\beta(t')\>_{eq,\,\th}$ is the equilibrium time correlation between the conjugate forces $\phi_\alpha$ and $\phi_\beta$
and where
we have written $\lam(t) \equiv \lam(\th(t))$ for shorthand.

Assuming the protocol speed is much slower than the timescale of system relaxation, by integrating by parts we find \cite{SivakCrooks12A}:
\be
  \<\del\phi_\alpha(t)\>_{\th(\tau)} \approx \f{\d\lam}{\d t}^\beta\til{g}^\lam_{\alpha\beta}
  \qquad
  \til{g}^\lam_{\alpha\beta}(\th) \equiv \int_{-\infty}^0\d t' C^{\th}_{\alpha\beta}(t')
\ee

\begin{figure*}
  \caption{\footnotesize
    An illustration of why the time derivative of the ensemble value of $\lambda$ is discontinuous.
    This system state is described by a variable $\theta$ with net drift $v$ and diffusion $D$ (explicitly $\dot{\theta}= v + \sqrt{2D}\eta$) and a control function $\lambda(\theta)$.
    Here we look at the ensemble $\Lambda$ of all possible system trajectories $\theta(\tau)$ such that $\theta(t_0)=\theta_0$ for a fixed point $(t_0, \theta_0)$.
    $(a-c)$ In blue is the probability of finding the system at a particular value of $\theta$ conditioned on the system being at $\theta_0$ at time $t_0$ for three different times: immediate before $t_0$ $(a)$, at $t_0$ $(b)$, and directly after $t_0$ $(c)$.
    Since the value of $\theta$ at $t_0$ is known, figure $(b)$ is a Dirac delta function.
    Figures $(a)$ and $(c)$ are identical Gaussians shifted forward and backward by $v\d t$.
    In red is the value of the control parameter $\lambda$ as a function of $\theta$.
    The important quantity is $\<\lambda(\theta)\>$ the value of $\lambda(\theta)$ at a given time averaged over trajectories in $\Lambda$.
    Note that while $\<\theta\>$ increases as a function of time (from $(a)$ to $(c)$) and $\lambda(\theta)$ increases as a function of $\theta$, it is not the case that $\<\lambda(\theta)\>$ increases as a function of time.
    This is because of diffusion.
    Between $(b)$ and $(c)$, the distribution of $\theta$ diffuses away from $\theta_0$.
    Because $\lambda$ is convex, this causes $\<\lambda(\theta)\>$ to increase.
    The same argument applies backwards in time as well.
    As we move backward in time from $(b)$ to $(a)$ the $\theta$ distribution also diffuses away from $\theta_0$, which in turn increases $\<\lambda(\theta)\>$ due to the convexity of $\lambda(\theta)$.
    This means that the minimum value of $\<\lambda(\theta)\>$ occurs at the intermediate time $t_0$.
    This is the origin of the discontinuity shown in $(d)$.
    $(d)$ In purple is a plot of $\<\lambda(\theta)\>$ over time averaged over the same ensemble of trajectories $\Lambda$.
    The time derivative has a discontinuity at $t=t_0$ related to $\d^2\lambda/\d\theta^2$ (the convexity of $\lambda$).
    The left and right sided limits of the time derivative (shown in blue and red) are the derivatives given by the Ito and reverse-Ito conventions for stochastic calculus~\cite{Gardiner04}.
  }
  \includegraphics{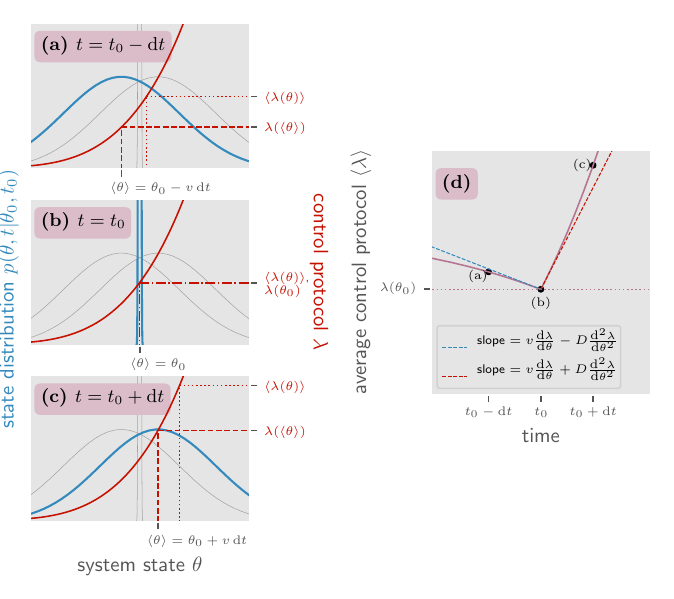}
  \label{fig:ito_velocities}
  \end{figure*}

We now have to extend this result to an ensemble of {\it stochastic} protocols which are all located at the same point $\th_0$ at time $t_0$ (to emphasize that we are referring to a specific point, we have switched from using the generic $\th$ and $t$ to $\th_0$ and $t_0$).
This process is a bit more challenging (see \rfeq{si:lin-resp-start}-\rfeq{si:lin-resp-end} in the Supplement for the exact details).
However the form of the result is manifestly the same: we only need to replace $\d\lam^\beta/\d t$ with the ensemble average of $\d\lam^\beta/\d t$, where the ensemble is taken over all control parameter trajectories $\th(\tau)$ such that $\th(t_0) = \th_0$.
As illustrated in Fig \ref{fig:ito_velocities}, this quantity is discontinuous at time $t_0$.
Its left and right-sided limits are:
\be
  \label{eq:lambda_derivatives}
  \l\<\f{\d\lam^\beta}{\d t^\pm}\hspace*{-2pt}\r\>_{\hspace*{-4pt}\th_0}\hspace*{-4pt}
  \equiv
  \lim_{t\to 0^\pm}
  \f{\d}{\d t'}\<\lam^\beta(t_0+t')|\th_0,t_0\>
  \approx
  v\f{\d\lam^\beta}{\d\th} \pm D\f{\d^2\lam^\beta}{\d\th^2}
\ee
which correspond to ensemble velocities under the reverse-Ito and Ito conventions~\cite{Gardiner04}.
Here we have dropped terms quadratic in $v$ and $D$.
Since the domain of \rfeq{eq:basic_linear_response} is $t'<0$, it the left-sided limit that is relevant in our calculation, and thus we obtain:
\be
  \label{eq:linear_response_result}
  \<\del\phi\>_\th
  \approx
  \l\<\f{\d\lam}{\d t^-}\r\>_{\hspace*{-4pt}\th}
  \til{g}^\lam
  \approx
  \l(v\f{\d\lam}{\d\th}(\th)-D\f{\d^2\lam}{\d\th^2}(\th)\r)\til{g}^\lam
\ee
To obtain this result we have required that both the control parameter velocity and the diffusion rate are small with respect to the system relaxation timescale at equilibrium, $\tau$.
Explicitly, we demand:
\begin{align}
  \label{eq:linear_response_approximations}
  v\tau/L \ll  1
  &&
  D\tau/L^2 \ll  1
\end{align}
where $L$ is related to the length scale associated with the control function $\lam(\th)$.
Finally, by using the result given by \rfequation{eq:linear_response_result} in \rfequation{eq:dissipation_no_linear_response}, we obtain our central result, \rfequation{eq:mainresult}.

\section*{Discussion}
The first term in \rfeq{eq:mainresult} represents the frictional dissipation arising from pushing the system out of equilibrium.
This term is a generalization of the dissipation rate found in the deterministically controlled coupled system studied in Ref.~\cite{SivakCrooks12A}.
To obtain that result, the authors made the assumption that the system moves much slower than the relaxation timescale of the system.
Concretely, this is the assumption that $v\tau/L \ll 1$.
In moving from \rfeq{eq:basic_linear_response} to \rfeq{eq:linear_response_result}, we make essentially the same assumption, but we also require that $D\tau/L^2\ll 1$.
This is merely the statement that the random movements of the control parameter cannot be large enough to take us away from the linear response regime.
Since we are searching for a {\it lower} energetic bound, this is certainly true in the limiting behavior.
The system of Ref.~\cite{SivakCrooks12A} can be seen as a special case where $D\to 0$ and the $v^2/D$ term is ignored.

The other two terms in \rfeq{eq:mainresult} are analogous to dissipation terms found in an earlier paper~\cite{Machta15} investigating a discrete system.
The first of these, $v^2/D$, is the energy required to break time reversal symmetry in the control parameter, i.e. the energy required for ``constantcy'' in the control clock~\cite{PietzonkaSeifert18,CaoTu15}.
The final term $Dg^\th$ can be thought of the energetic cost of straying from the optimal protocol, a geodesic~\cite{Crooks07}.

Figure \ref{fig:energy_curves}a plots the contribution to the total energy cost of a control protocol from the two sources of dissipation: the nonconservative work done by the control parameter on the $\bb x$ system ($W^{NC}=W^{on}-\Del F$ shown in blue) and the work required to break time reversal symmetry and direct the movement of $\th$ forward in time ($W^{TR}=\int \f{v^2}{D}\d t$ shown in red).
We see that they are minimized on opposite ends of the stochasticity spectrum.

In \rfequation{eq:mainresult} there is an implicit energetic tradeoff.
A control protocol that is very precise ($D\ll v$) (e.g. the red curve in Figure \ref{fig:energy_curves}b) may minimize the dissipation due to thermodynamic friction done by the control parameter on the system, however, such a protocol pays a high energetic cost for strongly breaking time-reversal symmetry.
On the other hand, a control protocol that only weakly breaks time-reversal symmetry ($D\gg v$) (e.g. the blue curve in Figure \ref{fig:energy_curves}b), pays a high energetic cost for undergoing suboptimal trajectories and performing redundant thermodynamic transitions.
Ref.~\cite{SivakCrooks12A} investigated the energetically optimal control {\it path} $\lam(t)$.
This work shows that there is also the question of the energetically optimal ``diffusive tuning'' between $v$ and $D$ which minimizes this tradeoff (e.g. the purple mark and curve in Figure \ref{fig:energy_curves}a, b).
In particular, the optimal control protocol is not deterministic $(D\neq 0)$.

As an example, consider a two-dimensional harmonic oscillator where the center is moved in a circle of radius $A$:
\be
  U(\th,\bb x) = \f{1}{2}k(\bb x - A\bb\lam)^2
  \qquad
  \bb\lam = (\cos\th,\sin\th)
\ee
Here $\bb\lam(\th) = (\cos\th,\sin\th)$ is the control function and $\phi(\bb x)=-kA\bb x$ is the conjugate force.
The Fisher information metric is given by $g^\lam = kA^2\bb I$ where $\bb I$ is the two-dimensional identity matrix and in the $\th$ basis we have $g^\th=\f{\d\lam^x}{\d\th}g^\lam_{xx}\f{\d\lam^x}{\d\th}+\f{\d\lam^y}{\d\th}g^\lam_{yy}\f{\d\lam^y}{\d\th}=kA^2$.
Using \rfeq{eq:mainresult}, we find the average dissipation rate:
\be
  \<\SProdRate\>
  =
  \f{v^2}{D}(1+g^\th D\tau) + Dg^\th(1-D\tau)
\ee
If we require that the protocol takes an average of $\Del t$ time per cycle, this fixes the net drift velocity $v=2\pi/\Del t$.
To find the minimum dissipation, we then optimize $S$ with respect to $D$.
This yields:
\be
  D^{(opt)} = \f{v}{\sqrt{g^\th}}(1+\epsilon)
\ee
where $|\epsilon|\ll 1$ is a small order correction due to the small $D\tau$ term.
Plugging this in yields a total dissipation per cycle of:
\be
  \<\SProd\>
  \geq 2\ThLen + \f{\til{\ThLen}^2}{\Del t}\l(1 - \f{(1+\epsilon)^2}{g^\th}\r)
\ee
where $\ThLen=2\pi\sqrt{g^\th}$ and $\til{\ThLen}=2\pi\sqrt{g^\th\tau}$ are the thermodynamic lengths of paths under the metrics $g^\lam$ and $\til{g}^\lam$.
In particular, we note that the dissipation remains bounded by $2\ThLen$ in the limit of an infinitely long protocol $\Del t\to \infty$.

In units where $\bt\neq 1$, the Fisher information metric is $g^\th = kA^2\bt$.
Thus the correction to the Sivak and Crooks bound can be neglected whenever the energy scale of the control is greater than the average thermal fluctuation.

The non-vanishing bound $2\ThLen$ scales with $\sqrt{g^\lam}$, the size of an average fluctuation in the system.
Thus it is subextensive and disappears in the macroscopic limit.
Its contribution is also dwarfed by thermodynamic friction when the control protocol is fast.
Therefore it is expected that this bound should only become relevant in slow microscopic systems.

We also note that this analysis only applies to {\it autonomous} thermodynamic machines: those whose control is independent of environmental signals.
In cases where the source of time-reversal symmetry breaking occurs externally,
e.g. a system reacting to cyclic changes in heat from the sun, this bound does not necessarily apply, though other bounds likely do.
This is because the origin of time-reversal symmetry breaking in such networks is environmental and thus no energy must be expended to drive the system in a particular direction.
Such {\it reactive} systems would be more appropriately characterized by Ref.~\cite{LargeSivak18}, which likewise addressed the question of optimality and energetic bounds in stochastically controlled systems, only without taking into account the cost of breaking time reversal symmetry.
Those authors found a minimum bound on the energy of control which cannot be made arbitrarily small in the presence of a noisy control protocol.
However, their bound is proportional to the magnitude of the control noise, and thus can be made arbitrarily small in the limit of noiseless protocols.

This work elucidates new constraints in the design of optimal thermodynamic machines.
In previous studies, it has been found that systems must dissipate more energy to increase the accuracy of their output~\cite{Barato15Thermodynamic,Horowitz17,Gingrich16,LanTu12}.
The optimal diffusive tuning found here indicates that below a certain level of accuracy, increasing precision actually {\it decreases} the dissipation cost.
In addition, the lower dissipation bound indicates that for very slow microscopic thermodynamic transformations, the dissipation cost no longer scales inversely with time.
The diminishing energetic returns from increasing the length of control protocols perhaps sets a characteristic timescale for optimal microscopic machines without time constraints.

\section*{Acknowledgements}
Thanks to Nicholas Read and Daniel Seara for useful discussions about our project.
This work was supported in part by a Simons Investigator award in MMLS (B.B.M.) and by NSF DMR-1905621 (S.J.B. and B.B.M.).

\bibliography{paper_bibliography}
\bibliographystyle{apsrev4-1}

\appendix

\section*{Supplemental Material}
\subsection*{Main Derivation}
Here we derive the main results (eq. 5 to eq. 14 in the main text) in greater detail.
The system's energy is given by:
\be
  U(\bb y) = \lam^\al(\th)\phi_\al(\bb x)
  \label{si:main-derivation-start}
\ee
Here $\bb\lam$, $\bb\phi$ are indexed by $\al$,$\bt$,$\ga$ and $\bb x$ is indexed by $i$,$j$,$k$.
We use natural units and unit temperature so that $\bt = T = 1$ and $\mu = D = \ga^{-1}$,
where $\mu$ is the mobility, $D$ is the diffusion, and $\ga$ is the resistance.
When $\mu$, $D$, $\ga$ and things like probability currents $j$ appear without indices, they refer to $\th$. In particular we use $\p_i$ and $\p_j$ as shorthand for $\p_{x^i}$ and $\p_{x^j}$. When they appear with roman indices ($i$,$j$,$k$), they refer to $\bb x$.
The Langevin equations are:
\begin{align}
  \begin{aligned}
    \dot{\th} = v + \sqrt{2D}\eta
    &&
    \dot{x}^i = D^{ij} F_j(\bb y) + \sqrt{2D^{ij}}\xi_j
    \\
    \<\eta(t)\eta(t')\>=\del(t-t')
    &&
    \<\xi_i(t)\xi_j(t')\>=\del_{ij}\del(t-t')
  \end{aligned}
\end{align}
where $\eta$ and $\xi$ are white noise functions and $F_j(\bb y) = -\p_j U(\bb y)$.
We denote $p(\bb y,t)$ to mean the probability of finding the system at state $\bb y = (\th,\bb x)$ at time $t$.
The probability currents are given by:
\be
  \begin{split}
  j(\bb y,t) &= \l[v - D\p_\th\r]p(\bb y, t)
    \\ &= p(\bb y, t)D\l[v\ga - \p_\th\log p(\bb y, t)\r]
  \\
  j^i(\bb y,t) &= D^{ij}\l[F_j(\bb y) - \p_j\r]p(\bb y, t)
    \\ &= p(\bb y, t)D^{ij}\l[F_j(\bb y) - \p_j\log p(\bb y, t)\r]
  \end{split}
\ee
The total averaged entropy production rate is the sum of the production rate for the two individual variables:
\begin{align}
  \<\SProdRate(t)\>
    &= \<\SProdRate^x+\SProdRate^\th\>
  \\&=
  \int\d\bb y \l[
    \f{j^i(\bb y,t)\ga_{ij}j^j(\bb y,t)}{p(\bb y, t)}
    +
    \f{j(\bb y,t)\ga j(\bb y,t)}{p(\bb y, t)}
  \r]
\end{align}
Starting with the $\th$ contribution:
\be
  \begin{split}
  \<\SProdRate^\th\>
  &= \int\d\bb y\f{j(\bb y,t)\,\ga\,j(\bb y,t)}{p(\bb y, t)}
  \\
  &= \int\d\bb y\,\l[v\ga - \p_\th\log p(\bb y, t)\r] j(\bb y,t)
  \\
  &=
  v\ga\int\d\bb y\, j(\bb y, t) - \int\d\bb y\,j(\bb y, t)\p_\th\log p(\bb y, t)
  \\
  &=
  v^2\ga - v\int\d\bb y\, p(\bb y, t)\,\p_\th\log p(\bb y, t)
  \\&\phantom{=}- \int\d\bb y\,j(\bb y, t)\,\p_\th\log p(\bb y, t)
  \end{split}
\ee
Since $p(\bb y, t)\p_\th\log p(\bb y, t) = \p_\th p(\bb y, t)$ and $\th$ is periodic, after integrating over $\th$, the middle term disappears leaving:
\be
  \<\SProdRate^\th\>
  = \f{v^2}{D} - \int\d\bb y\,j(\bb y, t)\p_\th\log p(\bb y, t)
\ee
Next we consider the $\bb x$ contribution:
\be
  \begin{split}
  \<\SProdRate^{x}\>
  &=
  \int\d\bb y\, \f{j^i(\bb y, t)\,\ga_{ij}\,j^j(\bb y, t)}{p(\bb y, t)}
  \\
  &=
  \int\d\bb y\, j^i(\bb y, t)(F_i(\bb y) - \p_i\log p(\bb y, t))
  \\
  &=
  -\int\d\bb y\,j^i(\bb y, t)\p_i U(\bb y)
  \\
  &\phantom{=}- \int\d\bb y\, j^i(\bb y, t)\p_i\log p(\bb y, t)
  \end{split}
\ee
Giving us:
\be
  \<\SProdRate\>
  =
  \f{v^2}{D} - \int\d\bb y\,\l[
    (j\p_\th + j^i\p_i)\log p + j^i\p_i U
  \r]
\ee
We can rewrite $U = \FE - \log p_{eq}$ where $\FE(\th)$ is the free energy for the system for a fixed $\th$ and $p_{eq}(\bb x|\th)=e^{\FE(\th)-U(\th, \bb x)}$ is the equilibrium Boltzmann distribution for fixed $\th$.
Then $\p_i U = -\p_i\log p_{eq}$ since $\FE$ is a function of $\th$ only:
\be
  \<\SProdRate\>
  =
  \f{v^2}{D} - \int\d\bb y\,\l[
    (j\p_\th + j^i\p_i)\log p - j^i\p_i\log p_{eq}(\bb x|\th)
  \r]
\ee
Now we may integrate by parts on each of the terms in the integral:
\be
  \<\SProdRate\>
  =
  \f{v^2}{D}
  +
  \int\d\bb y\,\l[(\p_\th j + \p_i j^i)\log p -(\p_i j^i \log p_{eq}(\bb x|\th) \r]
\ee
where the boundary terms disappear because $p(\bb y,t)$ goes to zero at the boundaries of $x$.

Now we assume that the system reaches a steady state so all time dependences drop out of the problem. In particular, the Fokker-Plank equation yields $\p_\th j(\bb y)+\p_i j^i(\bb y)=-\p_t p =0$, allowing us to remove the first term in the integral and swap $\p_ij^i(\bb y)=-\p_\th j(\bb y)$ in the second:
\be
  \<\SProdRate\> = \f{v^2}{D} + \int\d\bb y\, \l[\p_\th j(\bb y)\r] \log p_{eq}(\bb x|\th)
\ee
Next we expand $\p_\th j(\bb y)=v\p_\th p(\bb y) - D\p_\th^2 p(\bb y)$ and $\log p_{eq}(\bb x|\th)=\FE(\th) - U(\bb y)$, giving:
\be
  \<\SProdRate\> = \f{v^2}{D}
  +
  \int\d\bb y\,\l[v\p_\th p(\bb y) - D\p_\th^2 p(\bb y)\r](\FE(\th) -U(\bb y))
\ee
Now we integrate by parts to move the $\p_\th$, $\p_\th^2$ to the other term. Since $\th$ is periodic, we can always neglect the boundary terms:
\be
  \<\SProdRate\> = \f{v^2}{D}
  - \int\d\bb y\, p(\bb y) \l[v\p_\th+D\p_\th^2\r]\l[\FE(\th) - U(\bb y)\r]
\ee
We now need to compute the derivatives of $U$ and $\FE$ with respect to $\th$.
Recall we denote $\<\cdots\>_{\th}$ as an average over all microstates $\bb x$ with non-equilibrium weight $p(\bb x|\th)$ for a fixed value of $\th$. We also denote $\<\cdots\>_{\eq,\th}$ to indicate an average over all $\bb x$ with equilibrium Boltzmann weight $p_{\eq}(\bb x|\th)$ for a fixed value of $\th$:
\be
  \p_\th U(\th,\bb x) = \f{\p U}{\p\lam^\al}\f{\p\lam^\al}{\p\th}
  =
  \phi_\al(\bb x)\f{\p\lam^\al}{\p\th}(\th)
\ee
\be
  \p_\th^2 U(\th,\bb x)
  =
  \phi_\al(\bb x)\f{\p^2\lam^\al}{\p\th^2}(\th)
\ee
\be
  \p_\th\FE(\th) = \f{\p\FE}{\p\lam^\al}\f{\p\lam^\al}{\p\th}
  =
  \<\phi_\al\>_{\eq,\th}\f{\p\lam^\al}{\p\th}(\th)
\ee
\be
  \p_\th^2\FE(\th) = \<\phi_\al\>_{\eq,\th}\f{\p^2\lam^\al}{\p\th^2}(\th)
  -
  g^\lam_{\al\bt}\f{\p\lam^\al}{\p\th}\f{\p\lam^\bt}{\p\th}
\ee
where $g^\lambda_{\al\bt} = -\f{\p}{\p\lambda^\bt}\<\phi_\al\>_{\eq,\th}$ is the equilibrium Fisher information metric in the $\lambda$ basis.
Plugging these in gives:
\be
  \SProdRate
  =\f{v^2}{D}
  + \l[v\f{\p\lam^\al}{\p\th} + D\f{\p^2\lam^\al}{\p\th^2}\r]
    \del\phi_\al
  +Dg^\lam_{\al\bt}\f{\p\lam^\al}{\p\th}\f{\p\lam^\bt}{\p\th}
\ee
Where $\del\phi_\al = \phi_\al(\bb x)-\<\phi_\al\>_{\eq,\th}$ is the deviation of $\phi$ from its equilibrium value for fixed system state $\th$.
If we identify $g^\th \equiv \f{\p\lam^\al}{\p\th}g^\lam_{\al\bt}\f{\p\lam^\bt}{\p\th}$
as the Fisher information metric on $\th$ inherited from $\lam$, we get:
\be
  \SProdRate(\bb y)
  =
  \f{v^2}{D}
  + \l[v\f{\p\lam^\al}{\p\th} + D\f{\p^2\lam^\al}{\p\th^2}\r]
    \del\phi_\al
  +Dg^\th
\ee
To find the average dissipation rate when the system is at the point $\th$, we average over all $\bb x$ with weight $p(x|\th)$. This immediately yields the multidimensional analog of $(13)$:
\be
  \label{si:main-derivation-end}
  \<\SProdRate\>_\th
  =
  \f{v^2}{D}
  +Dg^\th(\th)
  +\l[v\f{\p\lam^\al}{\p\th}(\th) + D\f{\p^2\lam^\al}{\p\th^2}(\th)\r]
  \<\del\phi_\al\>_\th
\ee
Again, this is an exact expression. Next we use linear response to approximate $\<\del \phi_\alpha\>_\th$ assuming that $v$ and $D$ are small.

\subsection*{Linear Response Approximation}
Our starting point is the following expression~\cite{Zwanzig01} which gives the average linear response of $\phi$ at time $t_0$ to a specific control trajectory $\th(\tau)$:
\be
  \label{si:lin-resp-start}
  \<\del\phi_\al(t_0)\>_{\th(\tau)}
  =
  \int_{-\infty}^0
  \hspace*{-3pt}\d t'\l[\f{\d C^{\th(t_0)}_{\al\bt}}{\d t'}\r]\l[\bb\lam(t_0)-\bb\lam(t_0+t')\r]^\bt
\ee
with the multidimensional autocorrelation function for $\bb\phi$:
\begin{align}
  C_{\al\bt}^{\th(t_0)}(t)
  &= \<\del\phi_\al(0)\del\phi_\bt(t)\>_{\eq,\th(t_0)}
  \\
  &=
  \l\<\phi_\al(0)\phi_\bt(t_0)\r\>_{\eq,\th(t_0)}
  \\
  &\phantom{=}-
  \<\phi_\al(0)\>_{\eq,\th(t_0)}\<\phi_\bt(t)\>_{\eq,\th(t_0)}
\end{align}
This is the linear response function to a {\it single} path.

The expression we need is $\<\del\phi_\al(t_0)\>_{\th_0}$, the average of the above expression over {\it all paths} $\th(\tau)$ such that $\th(t_0)=\th_0$ for a specific point $(\th_0, t_0)$. The important point is that only $\bb\lam(t+t')$ is trajectory dependent, the other parts of the expression only depend on the value of the trajectory at the moment $t_0$.
Therefore we may write:
\be
  \<\del\phi_\al(t_0)\>_{\th_0}
  =
  \int_{-\infty}^0\d t'
  \l[\f{\d C^{\th_0}_{\al\bt}}{\d t'}\r]
  \<\lam^\bt(t_0)-\lam^\bt(t_0+t')|\th_0,t_0\>
\ee
where here $\<\cdots|\th_0,t_0\>$ represents an average over all possible control paths $\th(\tau)$ such that $\th(t_0)=\th_0$ weighted by their probability as given by the Langevin dynamics of $\th$.
By integrating by parts we are left with:
\be\label{eq:sup_linear_response_main}
  \<\del\phi_\al(t_0)\>_{\th_0}
  =
  \int_{-\infty}^0\d t' C^{\th_0}_{\al\bt}(t')
  \f{\d}{\d t'}\<\lam^\bt(t_0+t')|\th_0,t_0\>
\ee
To compute this, we need an expression for the expectation value of $\lam^\bt$ for this set of trajectories, which we may formally write as:
\be\label{eq:sup_lambda_expectation_value}
  \<\lam^\bt(t_0+t')|\th_0,t_0\>
  =
  \int_{-\infty}^\infty \d\th'\, p(\th_0+\th',t_0+t'|\th_0,t_0) \lam^\bt(\th_0+\th')
\ee
Note that despite the fact that $\th$ is periodic, the integration bounds here are not. This is because $\th'=2\pi$ in this context refers to the control parameter making a full cycle in time $t'$ which is not the same as it not moving ($\th'=0$).

Since the stochasticity of $\th$ is driven by Gaussian noise, it's trivial to write down $p$:
\be
  p(\th_0+\th',t_0+t'|\th_0,t_0) = \f{1}{\sqrt{4\pi D|t'|}}
  e^{-\f{(\th'-vt')^2}{4D|t'|}}
\ee
which is a Gaussian that diffuses away from a Dirac delta function as $|t'|>0$ (see Fig 2 from the main text).
We also Taylor expand $\lam^\bt(\th_0+\th')$ about $\th_0$:
\be
  \lam^\bt(\th_0+\th')
  =
  \lam^\bt(\th_0)+\th'\p_\th\lam^\bt(\th_0)
  +
  \f{1}{2!}(\th')^2\p_\th^2\lam^\bt(\th_0)
  +\cdots
\ee
Putting these together gives:
\be
  \<\lam^\bt(t_0+t')|\th_0,t_0\>
  =
  \sum_{k=0}\f{\p_\th^k\lam^\bt(\th_0)}{k!\sqrt{4\pi D|t'|}}
  \int_{-\infty}^{\infty}\d\th'\,e^{-\f{(\th'-vt')^2}{4D|t'|}}(\th')^k
\ee
Solving this integral for each value of $k$ will yield terms of the form:
\be
   (D|t'|)^m(vt')^n\p^{2m+n}_\th\lam^\bt(\th_0)
\ee
An underlying assumption of this work is that we are in the near-equilibrium regime where the perturbation on the system is weak. We treat $v$ and $D$ as small parameters and only keep terms of order $m+n=1$ in $v$, $D$:
\be\label{eq:sup_lambda_approx}
  \<\lam^\bt(t_0+t')|\th_0,t_0\>
  =
  \lam^\bt(\th_0)
  +
  vt'\p_\th\lam^\bt(\th_0)
  +
  D|t'|\p^2_\th\lam^\bt(\th_0)
  +
  \Order(2)
\ee
We can see that to leading order, the time-derivative of this expression will be time-independent.
Thus, to leading order we can replace $\f{\d}{\d t'}\<\lam^\bt(t+t')|\th_0,t_0\>$ with its value at $t'=0$. However, due to the presence of the absolute value terms $|t'|$, the time derivative at $t'=0$ is discontinuous. This is more than a minor detail; it's connected with the choice of whether to define derivatives in stochastic calculus using left sided limits or right sided limits analogous to choices in Riemann sums~\cite{Gardiner04}.

We introduce the following notation to distinguish between the left and right-sided limits of the time derivative of the conditional expectation value of $\lambda$:
\be
  \l\<\f{\d\lam^\bt}{\d t^{\pm}}\r\>_{\hspace*{-3pt}\th_0}
  \equiv
  \lim_{t'\to 0^\pm}\f{\d}{\d t'}\<\lam^\bt(t_0+t')|\th_0,t_0\>
  \approx
  v\p_\th\lam^\bt(\th_0) \pm D\p^2_\th\lam^\bt(\th_0)
\ee
As one may expect, this is exactly the formula for the net drift of $\lambda^\bt$ at $t'=0$ given by the Ito formula under the Ito ($+$) and reverse-Ito ($-$) conventions~\cite{Gardiner04}.

Plugging this result into \rfeq{eq:sup_linear_response_main} and using the definition of thermodynamic friction:
\be
  \til{g}^\lam_{\al\bt} = \tau_{\al\bt}\circ g^{\lam}_{\al\bt} = \int_{-\infty}^0\d t'\,C^\th_{\al\bt}(t')
\ee
yields the time-independent expression for the average linear response:
\be
  \label{si:lin-resp-end}
  \<\del\phi_\al\>_{\th_0}
  =
  \til{g}^\lam_{\al\bt}(\th_0)\l\<\f{\d\lam^\bt}{\d t^-}\r\>_{\hspace*{-3pt}\th_0}
\ee
which in turn gives an average entropy production rate for the system when it's at the state $\th_0$:
\be
  \label{eq:sup_multivar_dissipation_form_2}
  \<\SProdRate\>_{\th_0}
  =
  \l\<\f{\d\lam^\al}{\d t^+}\r\>_{\hspace*{-3pt}\th_0}
    \til{g}^\lam_{\al\bt}(\th_0)
  \l\<\f{\d\lam^\bt}{\d t^-}\r\>_{\hspace*{-3pt}\th_0}
  +
  \f{v^2}{D}+Dg^\th(\th_0)
\ee
What about the higher order terms in $\rfeq{eq:sup_lambda_approx}$?
We can neglect them under the assumption that the speed of control is small compared to the excitation timescale $\tau$ of the system at equilibrium.
However, an explicit mathematical statement of this requirement is challenging because of the unknown form of $\lambda(\th)$.
Since $C$ is a decay function, we expect roughly that:
\be
  \int_{-\infty}^0\d t'\, (t')^k C^\th_{\al\bt}(t')  \sim g^\lam_{\al\bt} \tau^{k+1}
\ee
Thus in keeping terms of order $m+n>1$ we would generate additional contributions
to \rfeq{si:lin-resp-end} of the form:
\be
  (D\tau)^m(v\tau)^n \l[g^\lam_{\al\bt}\p^{2m+n}_\th\lam^\bt(\th)\r]
\ee
Comparing these to the leading order terms
$(D\tau)g^{\lam}_{\al\bt}\p_\th^2\lam^\bt$
and
$(v\tau)g^{\lam}_{\al\bt}\p_\th\lam^\bt$, we can see that for a reasonably behaved function $\lambda(\th)$ with a characteristic length scale $L$ we can summarize our assumption via the requirements:
\be
  \label{eq:sup_constraints} v\tau/L \ll 1\qquad\qquad D\tau/L^2 \ll 1
\ee
\vspace*{0.5em} 

Essentially this is the requirement that $\l\<\f{\d\lam^\al}{\d t}(t)\r\>$ (under the reverse-Ito convention) remains relatively constant over the system relaxation timescale $\tau$. This is the natural stochastic generalization of the constraint imposed by [3].

However, because we have not explicitly given a definition for $L$, this requirement is admittedly a little vague.
One major complication in doing so arises from the fact that because we only care about the {\it total} dissipation bounds, we only want to keep the terms in \rfeq{eq:sup_multivar_dissipation_form_2} that contribute to the leading order behavior of the {\it integral} of \rfeq{eq:sup_multivar_dissipation_form_2}.
Thus, while it may be the case that for a specific point $\th_0$, the second order term dominates: $(v\tau)^2\p_\th^2\lam(\th_0) \gg (v\tau)\p_\th\lam(\th_0)$, we still want to drop the higher order term because its contribution to the total integral is subleading.
The actual formal constraints dictating when this approximation is appropriate is further complicated by the unconstrained behavior of $\til{g}(\th)$ and $\lambda(\th)$.
However, it should be clear that as we approach equilibrium behavior ($D$, $v\to 0$), the kept terms dominate over the dropped terms.
We feel that the constraint given in \rfeq{eq:sup_constraints} satisfactorily captures this idea.

\end{document}